\documentclass[11pt,a4paper,fleqn]{article}   
\catcode`\@=11
\@addtoreset{equation}{section}
\def\theequation{\arabic{section}.\arabic{equation}}
\def\appendix{\renewcommand{\thesection}{\Alph{section}}\setcounter{section}{0}
              \renewcommand{\theequation}
            {\mbox{\Alph{section}.\arabic{equation}}}\setcounter{equation}{0}}

\def\maketitle{\thispagestyle{empty}\setcounter{page}0\newpage
                \renewcommand{\thefootnote}{\arabic{footnote}}
                  \setcounter{footnote}0}
\renewcommand{\thanks}[1]{\renewcommand{\thefootnote}{\fnsymbol{footnote}}
               \footnote{#1}\renewcommand{\thefootnote}{\arabic{footnote}}}

\renewcommand{\title}[1]{\begin{center}\Large\bf #1\end{center}\rm\par\bigskip}
\renewcommand{\author}[1]{\begin{center}\Large #1\end{center}}
\newcommand{\address}[1]{\begin{center}\large #1\end{center}}

\def\babs{\hrule\par\begin{description}\item{Abstract: }\it} 
\def\eabs{\par\end{description}\hrule\par\medskip\rm}
\renewcommand{\date}[1]{\par\bigskip\par\sl\hfill #1\par\medskip\par\rm}

\def\beq{\begin{eqnarray}}    
\def\be{\begin{eqnarray}}
\def\eeq{\end{eqnarray}}      
\def\ee{\end{eqnarrayn}}
\def\R{{\hbox{{\rm I}\kern-.2em\hbox{\rm R}}}}   
\def\H{{\hbox{{\rm I}\kern-.2em\hbox{\rm H}}}}   
\def\N{{\hbox{{\rm I}\kern-.2em\hbox{\rm N}}}}   
\def\C{{\ \hbox{{\rm I}\kern-.6em\hbox{\bf C}}}} 
\def\Z{{\hbox{{\rm Z}\kern-.4em\hbox{\rm Z}}}}   
\def\dir{/\kern-.7em D\,}                          

\def\be{\begin{equation}}
\def\ee{\end{equation}}
\def\bea{\begin{eqnarray}}
\def\eea{\end{eqnarray}}

\renewcommand{\title}[1]{\begin{center}\Large\bf #1\end{center}\rm\par\bigskip}
\renewcommand{\author}[1]{\begin{center}\Large #1\end{center}}

\begin{document}

\title{Minisuperspace Approach  of Generalized  Gravitational 
 Models  }
\author{ Guido Cognola\thanks{cognola@science.unitn.it}}
\address{
Dipartimento di Fisica, Universit\`a di Trento \\ 
and Istituto Nazionale di Fisica Nucleare \\ 
Gruppo Collegato di Trento, Italia}
\author{ Sergio Zerbini\thanks{zerbini@science.unitn.it}}
\address{
Dipartimento di Fisica, Universit\`a di Trento \\ 
and Istituto Nazionale di Fisica Nucleare \\ 
Gruppo Collegato di Trento, Italia}

\begin{abstract}
Motivated by the dark energy issue, the minisuperspace approach for 
 general relativistic cosmological  theories is outlined.

\end{abstract}


\section{Introduction}

It is well known that recently there has been found  strong 
 evidence for an accelerate expansion of our  universe, apparently due to the 
so called dark energy. With regard to this 
issue, here  we would like to make  some considerations 
involving general relativistic  theories of gravitation. 
 In fact, recently alternative and 
geometric descriptions for the dark energy in modern cosmology have been 
proposed and discussed in several related issues
\cite{turner,nojiri0}. Such models are higher derivative gravitational 
theories, thus they may contain instabilities  \cite{ins} and deviation from 
Newton gravity \cite{lue}. However, if one takes quantum effects into account,
one can get a viable theory \cite{NO1}. 
The Palatini method has also been applied in consistent way 
\cite{pala,alle1,alle2} and the evaluation of the black hole entropy 
within these models has been investigated in \cite{brevik}.

To this aim, we shall consider a  general relativistic theories,
(see for example \cite{barrow83,fara}), namely  let us assume 
that our model is described by the action
\be
\nonumber
S=\frac{1}{16 \pi G}\int d^4x \sqrt{-g} f(R) \,,
\label{genR}
\ee
with $f(R)$ depending only on the scalar curvature. 

As first example, let us consider the Lagrangian \cite{turner}
\be
\nonumber
f(R)=\left[ 
R-\frac{\mu^4}{R} \right]\,,
\label{invR}
\ee
where $\mu$ is a new cosmological parameter \cite{turner}. 
As is well known , there exist constant curvature   
de Sitter and AdS vacuum solutions  such that
\be
R^2_0=3\mu^4\,.
\label{ccv}
\ee

Another well known example,  is given by the choice 
\be
f(R)=R+\gamma R^2-2\Lambda\,,
\label{q}
\ee 
where the the other possible quadratic term giving by the Weyl invariant has
been omitted because is 
vanishing for space-time we are dealing with 
(see, for example \cite{barrow83}).

As a third example, let us consider an effective  Coleman-Weinberg like model  
\be
f(R)=R+ R^2(\gamma+\beta \ln (\frac{R}{\mu^2})\,,
\label{q1}
\ee 
where $\gamma$, $\beta$ and $\mu$ are suitable constants.

\section{Minisuperspace approach}

Our aim in this section will be  the issue of a minisuperspace Lagrangian 
description, in order to investigate classical and quantum aspects, like the 
stability and canonical quantization. For these reasons,
one has to  restrict  to FRW isotropic and homogeneous metrics with
constant  spatial section. We choose a spatial flat metric, namely 
\be
ds^2=a^2(\eta)\left(-N^2(\eta) d\eta^2+d^2 \vec{ x}\right)\,,
\ee
where $\eta$ is the conformal time, $a(\eta)$ the  cosmological 
factor and $N(\eta)$ an arbitrary lapse function, which describes the gauge 
freedom associated with the reparametrization invariance of the minisuperspace 
gravitational model. For the above metric, the scalar curvature reads
\be
R=6\left(\frac{a''}{a^3 N^2}-\frac{a' N'}{a^3 N^3}\right)\,,   
\ee
in which $'$ stands for $\frac{d}{d \eta}$.
 
If one plugs this expression in the Eq. (\ref{genR}), one obtains, 
 an higher derivative Lagrangian theory. Higher derivative Lagrangian theory
may be treated canonically by means the Ostrogadski method 
(see, for example \cite{pagani88} and references cited therein).

Here we have found more convenient to follow the method outlined in ref. 
\cite{vilenkin85}. To deal with a non standard  higher derivatives Lagragian 
system, we  make use of a 
Lagrangian multiplier $\lambda$  and we  write 
\be
S=\frac{V_3}{16 \pi G} \int d\eta N a^4 \left[ L_0(R)-
\lambda \left( R-6(\frac{a''}{a^3 N^2}-\frac{a' N'}{a^3 N^3})\right)\right]\,.
\ee
Making the variation with respect to $R$, one gets
\be
\lambda=\frac{d f(R)}{d R}\,.
\ee
Thus, substituting this value and making a standard integration by part, one 
arrives at the Lagrangian, which will be our starting point
\be
L(a,a',R,R')=-6\frac{a'^2}{N}\frac{d f(R)}{d R}-6\frac{a' a R'}{N}
\frac{d^2 f(R)}{d^2 R}+
Na^4 (f(R)-R\frac{d f(R)}{d R})\,.
\label{l}
\ee
It should be noted that $N$ appears as ``einbein'' Lagrangian multiplier, 
as it should be,
 reflecting the parametrization invariance of the action. In fact the 
Lagragian  is quasi-invariant with respect to the infinitesimal gauge 
transformation
\be
\delta a=\epsilon(\tau)a'(\tau)\,,\,\,\,\,
\delta R=\epsilon(\tau)R'(\tau)\,,\,\,\,\,
\delta N=\frac{d}{d \tau}\left[\epsilon(\tau)N(\tau)\right]\,.
\ee
As a consequence,  we have the (energy) constraint 

\be
\frac{\partial L}{\partial N}=0\,,
\ee
namely
\be
6\frac{a'^2}{N^2}\frac{d f(R)}{d R}+6\frac{a' a R'}{N^2}\frac{d^2 f(R)}{d^2 R}-
a^4 (f(R)-R\frac{d f(R)}{d R})=0\,,
\label{en1}
\ee
and we may choose, for example,  the gauge $N=1$.
The other Eqs. of motion are
\be
\frac{d}{d \eta}
\left[\frac{2a'}{N}\frac{d f(R)}{d R}+\frac{a R'}{N}
\frac{d^2 f(R)}{d^2 R}\right]
=\frac{a' R'}{N}\frac{d^2 f(R)}{d^2 R}-\frac{2}{3}a^3
(f(R)-R\frac{d f(R)}{d R})\,,
\ee
\be
\frac{d}{d \eta}
\left[\frac{a a'}{N}\frac{d^2 f(R)}{d^2 R}\right]
=\frac{RNa^4}{6}\frac{d^2 f(R)}{d^2 R}+\frac{a'^2}{N}\frac{d^2 f(R)}{d^2 R}+
\frac{a' R'a}{N}\frac{d^3 f(R)}{d^3 R}\,.
\ee
The conserved quantity is the energy, computed with the standard Legendre 
transformation
\be
E=-6\frac{a'^2}{N}\frac{d f(R)}{d R}-6\frac{a' a R'}{N}\frac{d^2 f(R)}{d^2 R}-
Na^4 (f(R)-R\frac{d f(R)}{d R})\,.
\label{en}
\ee
$E$ is vanishing on shell due to the Eq. of motion for the einbein $N$.

We shall be  interested in models which admit solution with constant 
4-dimensional curvature of the de Sitter type, namely
\be
R=R_0\,,\,\,\,\,a_0=\frac{A}{\eta}\,,\,\,\,\, A^2R_0=12\,. 
\ee
If we plug this  particular solutions in the above Eqs. of motion,
we get the condition \cite{barrow83}
\be
2f(R_0)=R_0\frac{d f(R)}{d R}(R_0)\,,
\label{B}
\ee
which may be used to find  the constant curvature $R_0$.
For the  model defined by Eq. (\ref{invR}), 
Eq. (\ref{B}) leads again to  the condition 
\be
R_0^2=3\mu^4\,,
\ee
while for the Lagrangian (\ref{q}), Eq. (\ref{B}) gives 
\be
R_0=4\Lambda\,,
\ee
and for the Coleman-Weinberg like model gives
\be
R_0=\frac{1}{c_2}\,,\,\,\,\,\,\,\, R_0=0\,.
\ee 
It is easy to check that such kind of solutions are physically ones, because
we have
\be
E=\frac{6 A^2}{\eta^4}
\left[ 1-\frac{2f(R_0)}{R_0\frac{d f}{d R}(R_0)} \right]\,,
\ee
namely  they satisfy identically the energy constraint $E=0$. Thus, the 
condition  (\ref{B}) turns out to be a necessary and sufficient condition
in order to have physical constant curvature solutions.

In order to  investigate the Hamiltonian formalism, it is  
convenient to make the following change of variables  \cite{vilenkin85}: 
$N \rightarrow N$,
$ a \rightarrow q$ and $ R \rightarrow \phi$, defined by
\be
\frac{d f(R)}{d R}=B^2e^{2 \phi}\,
\ee
\be
a=q e^{-\phi}\,.
\ee
In the  first, $B$ is a suitable constant,  fixed by means of
\be
B^2=\frac{d f(R)}{d R}(R_0)\,,
\ee
and $R$, as a function of the 
new variable $\phi$, is defined 
implicitely  $R=R(\phi)$. 

For example, for the choice  (\ref{invR}), one has
\be
\frac{d f(R)}{d R}(R)=1+\frac{\mu^4}{R^2}=B^2e^{2 \phi}\,,
\ee
and
\be
R=\frac{\mu^2}{(B^2e^{2 \phi}-1)^{1/2}}\,.
\ee

For the Lagrangian (\ref{q}), one obtains
\be
R=\frac{B^2e^{2\phi}-1}{2\gamma}\,.
\ee
In the case of Coleman-Weinberg like model, one only has
\be
B^2e^{2\phi}=1+2R(\gamma+\beta \ln \frac{R}{\mu^2})+\beta R \,.
\ee
Thus,  it is not possible to obtain explicitly $R$ as a function of the 
new variable $\phi$.

The de Sitter like solution corresponds to
\be
\phi_0=0\,,\,\,\,\,\,\,q_0=\frac{A}{\eta}\,,\,\,\,\,\,\, A^2R_0=12\,.
\label{ds}
\ee
A direct calculation leads to Lagrangian
\be
L=\frac{6}{N}\left[(2q-q^2)^2 \phi'^2-q'^2 \right]-N V(\phi,q)\,,
\label{newl}
\ee
in which  the potential reads
\be
V(\phi,q)=-\frac{q^4}{B^2}e^{-4\phi}
\left[f(R(\phi))- R \frac{d f(R)}{dR}(\phi)\right]\,.
\ee
For example, for the Lagrangian (\ref{invR}), we have 
(see also\cite{turner,nojiri0,ins})
\be
V(\phi,q)=\frac{q^4 e^{-4\phi}}{B^2}\left[ B^2e^{2\phi}-1\right]^{1/2}\,,
\ee
while for the Lagrangian (\ref{q}), one obtains
\be
V(\phi,q)=\frac{q^4}{B^2}e^{-4\phi}\left[2\Lambda+
\frac{\left(B^2e^{2\phi}-1\right)^2}{4\gamma} \right]\,,
\ee
The conserved energy reads
\be
E= \frac{6}{N}\left[(2q-q^2)^2\phi'^2-q'^2 \right]+N V(\phi,q)\,,
\ee
and it is vanishing on shell. The corresponding Hamiltonian is
\be
H=N\left[\frac{1}{24(2q-q^2)}P^2_{\phi}-\frac{1}{24}P^2_q+V(\phi,q)\right]\,.
\label{ham}
\ee

The Lagrangian equation of motion, in the gauge  $N=1$, read
\be
q''+(1-q)\phi'^2+\frac{q^3}{3}\frac{e^{-4\phi}}{B^2}
\left[f(R)-R\frac{df(R)}{dR}\right]=0\,,\label{q2}
\ee
\be
(2q-q^2)\phi''+2(q'-qq')\phi'-\frac{q^4}{6B^2}e^{-4\phi} 
\left[-2 f(R)+R\frac{df(R)}{dR}\right]=0\,.
\label{r1}
\ee
If the Lagrangian satisfies the condition (\ref{B}), it is easy to check
that the one has again the de Sitter solution (\ref{ds}). 

We conclude this Section writing down the equations for the small 
disturbances around the de Sitter solution, namely
\be
q=q_0+\delta q\,,\,\,\,\,\,\,\phi=\delta \phi\,.
\ee
Taking Eq. (\ref{B}) into account again, the equations for the small 
disturbance around de Sitter solution turn out to be
\be
\frac{d^2\delta q}{d^2 \eta}-\frac{6}{\eta^2} \delta q=0\,,
\label{dq}
\ee
\be
(2-\frac{A}{\eta})\frac{d^2\delta \phi}{d^2 \eta}-
\frac{2}{\eta}(1-\frac{A}{\eta})\frac{d\delta \phi}{d \eta}-
4 R_0\frac{A^3}{\eta^3}
 \left[1- \frac{2f(R_0)}{R_0^2\frac{d^2f}{d^2 R_0}}\right]
\delta \phi=0\,.
\label{dr}
\ee
Some remarks are in order. First the small disturbance equations are 
decoupled in the conformal time. Second,   
the equation associated with the variable $q$ is universal, namely it 
does not depend explicitly on function $f(R)$, but we remind that the constant value $R_0$ depends on it. 
 Third, the general  solution of the first 
equation is not hard to find and reads
\be
\delta q=c_1 \eta^3+c_2 \eta^{-2}\,.
\ee
However, the perturbed solution $q=q_0+\delta q$ must satisfy the energy 
constraint $E=0$  and this leads to 
\be
\delta E=0\,,
\ee
namely, around the de Sitter solution
\be
q_0'\delta q'-q_0'' \delta q=0\,.
\ee
As a result $c_1=0$ and we have
\be
q=\frac{A}{\eta}[1+\frac{\eta_0^2 \delta q_0}{A \eta}]\,.
\ee
Recall that the relation between the conformal time and the cosmological time
$t$ may be written 
\be
\eta=\eta_0 e^{\frac{t}{A}}\,,  
\ee
and $12A=R_0^2$,  $R_0$ being the de Sitter curvature and $t=0$ corresponds
to $\eta_0$.  

Thus, the solution remains small with respect the de Sitter one  for
\be
\eta > \frac{12 \eta_0^2 \delta q_0}{R_0^2}\,,
\ee   
$R_0$ being the de Sitter curvature.  

Along the same lines, one may investigate the Starobinski model 
\cite{staro} and its generalization
including the Brans-Dicke field \cite{bodo} and its brane-world 
generalization \cite{noji00}.

\section{Conclusions}

In this paper, we have presented a minisuperspace approach for 
general relativistic pure gravitational models.
The inclusion of the matter can be easily taken into account. 
A canonical approach has been presented by means of the methods 
of ref. \cite{vilenkin85}. These models are, in general, instable,
due to the presence, at the beginning, of higher derivative terms \cite{ins}.
However, the inclusion of quantum effects may resolve the problem \cite{NO1}.

\section*{Acknowledgments}
We would like to thank E. Elizalde, S. Nojiri and S.D. Odintsov for 
stimulating  collaboration.

\end{document}